\documentclass[letterpaper, 10 pt, conference]{ieeeconf}  

\IEEEoverridecommandlockouts                              

\usepackage{graphicx}
\usepackage{url}  

\title{\LARGE \bf
Robust Deep Learning Framework For Predicting \\
Respiratory Anomalies and Diseases
}

\author{Lam~Pham$^{1}$, 
             Ian~McLoughlin$^{2}$, 
             Huy~Phan$^{1}$, 
             Minh~Tran$^{3}$,
             Truc~Nguyen$^{4}$, 
             Ramaswamy~Palaniappan$^{1}$ %
\thanks{L. Pham, I. McLoughlin, H. Phan and R. Palaniappan are with the University of Kent, School of Computing, Medway, Kent, UK.}%
\thanks{I. McLoughlin is with the National Engineering Laboratory for Speech and Language Information Processing at the University of Science and Technology of China, Hefei, P.R.China.}%
\thanks{M. Tran is with Nuffield Department of Anaesthesia, University of Oxford, UK.}%
\thanks{T. Nguyen is with the Signal Processing and Speech Communication Lab, Graz University of Technology, Austria.}%
}

\begin{document}

\maketitle
\thispagestyle{empty}
\pagestyle{empty}

\begin{abstract}

This paper presents a robust deep learning framework developed to detect  respiratory diseases from recordings of respiratory sounds.
The complete detection process firstly involves front end feature extraction where recordings are transformed into spectrograms that convey both spectral and temporal information. Then a back-end deep learning model classifies the features into classes of respiratory disease or anomaly.
Experiments, conducted over the ICBHI benchmark dataset of respiratory sounds, evaluate the ability of the framework to classify sounds.
Two main contributions are made in this paper.
Firstly, we provide an extensive analysis of how factors such as respiratory cycle length, time resolution, and network architecture, affect final prediction accuracy.  Secondly, a novel deep learning based framework is proposed for detection of respiratory diseases and shown to perform extremely well compared to state of the art methods.
\newline

\indent \textit{Clinical relevance}--- Respiratory disease, wheezes, crackles, Convolutional neural network (CNN), recurrent neural network (RNN).
\end{abstract}

\section{INTRODUCTION}

According to the World Health Organization, respiratory illness is one of the most common mortality factors worldwide~\cite{who}. This includes diseases and conditions such as lung cancer, tuberculosis, asthma, chronic obstructive pulmonary disease (COPD), and lower respiratory tract infection (LRTI).
In most cases, an effective way to combat mortality is early detection -- helping to limit spread or increase effectiveness of treatment.
Detecting anomaly sounds such as \textit{Crackles} or \textit{Wheezes} during lung auscultation (listening to the sounds produced) is an important aspect of a medical examination to diagnose respiratory disease.
Both sounds are categorised into a group of adventitious sounds, which may indicate pulmonary disorders~\cite{sound_early_03}.
If automated methods can be developed to detect these anomaly sounds, it may be useful in enhancing the early detection of respiratory disease in future.
Automated analysis of respiratory sounds has a long history~\cite{sound_early_03}, however the research field attracted little attention until
recent years when robust machine learning techniques were developed.
Current machine learning approaches to respiratory sound analysis tend to rely upon frame-based representations. 
Most use Mel-frequency cepstral transformation~\cite{lung_hmm_01, lung_hmm_02} to derive feature vectors.
These vectors are then classified by traditional machine learning models such as Support Vector Machine~\cite{lung_svm_01}, Hidden Markov Model~\cite{lung_hmm_01, lung_hmm_02}, or decision trees~\cite{lung_tree_18}.
Meanwhile, in the related field of sound event detection, deep learning techniques were introduced which achieved strong and robust detection performance for general sounds~\cite{ivmCNNsounddet},~\cite{ivmearly_2018}.
In those systems, feature extraction involves generating two-dimensional spectrograms, able to represent both temporal and spectral information, and do so over a much wider time context than single frame analysis.
These methods, since introduced for lung sound analysis, feed spectrogram features into back-end classifiers which exploits powerful network architectures such as CNN~\cite{lung_cnn_01, lung_cnn_02} or RNN~\cite{lung_rnn_01, lung_rnn_02}. 
Although recent publications increasingly achieve good performance in terms of classification of respiratory sounds, it is hard to compare systems due to the use of different datasets, mainly collected by authors, and often not publicly available.
To alleviate this situation, the 2017 Internal Conference on Biomedical Health Informatics (ICBHI)~\cite{lung_dataset} provided a benchmark respiratory sound dataset, for which challenging tasks were clearly defined. \\
In this paper, we explore system performance on the ICBHI dataset.
In particular, we isolate the effect of various factors related to front-end feature extraction such as respiratory cycle length and time resolution, to determine how they affect detection performance.
Furthermore, we report extensive experiments to evaluate various deep learning architectures (mainly based on CNN and RNN), enabling us to  propose a novel high performance and robust deep learning framework for respiratory disease detection.
\section{ICBHI challenge and our setup}
The ICBHI challenge provided a large dataset of respiratory sounds, collected from a total of 128 patients over 5.5 hours.
The dataset comprises 920 audio recordings, each which contains one to four different types of cycles called \textit{Crackle}, \textit{Wheeze}, \textit{Crackle \& Wheeze}, and \textit{Normal} accompanied by onset and offfset time labels.
These cycles have various recording lengths (from 0.2\,s up to 16.2\,s), with the number of cycles being unbalanced, with 1864, 886, 506 and 3642 cycles respectively for \textit{Crackle}, \textit{Wheeze}, \textit{Crackle \& Wheeze}, and \textit{Normal}.
The full audio recordings have various lengths from 10 to 90\,s, and used a wide range of sampling frequencies from 4\,kHz to 44.1\,kHz. 
Each audio recording identifies the patient's situation in terms of being healthy or exhibiting one of the following respiratory diseases or conditions: COPD, Bronchiectasis, Asthma, Upper and Lower respiratory tract infection, Pneumonia, Bronchiolitis.
Given this metadata, the ICBHI challenge is separated into two main tasks.
Task 1, referred to as respiratory anomaly classification, is separated into two sub-tasks.
The first classifies four different types of cycle (\textit{Crackle}, \textit{Wheeze}, \textit{Crackle \& Wheeze}, and \textit{Normal}).
The second sub-task aims to classify into two groups of \textit{Normal} and \textit{Anomaly} cycles (the latter consisting of \textit{Crackle}, \textit{Wheeze}, \textit{Both Crackle \& Wheeze}). 
Task 2, referred to as respiratory disease prediction, also comprises two sub-tasks.
The first classifies three groups of disease conditions known as \textit{Healthy}, \textit{Chronic Disease} (i.e. COPD, Bronchiectasis and Asthma) and \textit{Non-Chronic Disease} (i.e. Upper and Lower respiratory tract infection, Pneumonia, and Bronchiolitis).
The classification in the second sub-task is for two groups of \textit{healthy} or \textit{unhealthy} (i.e. the \textit{chronic} and \textit{non-chronic} disease groups combined).
While Task 1 is evaluated over respiratory cycles, Task 2 is evaluated over entire audio recordings. 
%
%
In this paper, we attempt all of the ICBHI challenge tasks.
Firstly, we separate the ICBHI dataset (6898 respiratory cycles for Task 1 and 920 entire recordings for Task 2) into five-folds for cross validation.
We assess the front-end feature extraction factors (namely cycle length, time resolution and network architecture) over just the first fold, but
evaluate the best system configuration over all folds, comparing the mean performance to the state of the art. 
We follow the ICBHI criteria and settings, and give results in terms of sensitivity, specitivity and ICBHI score as defined in~\cite{lung_dataset, lung_rnn_01}.
%
%
\section{Proposed baseline system}
%
\begin{figure}[tb]
    \centering
    \includegraphics[width =1.0\linewidth]{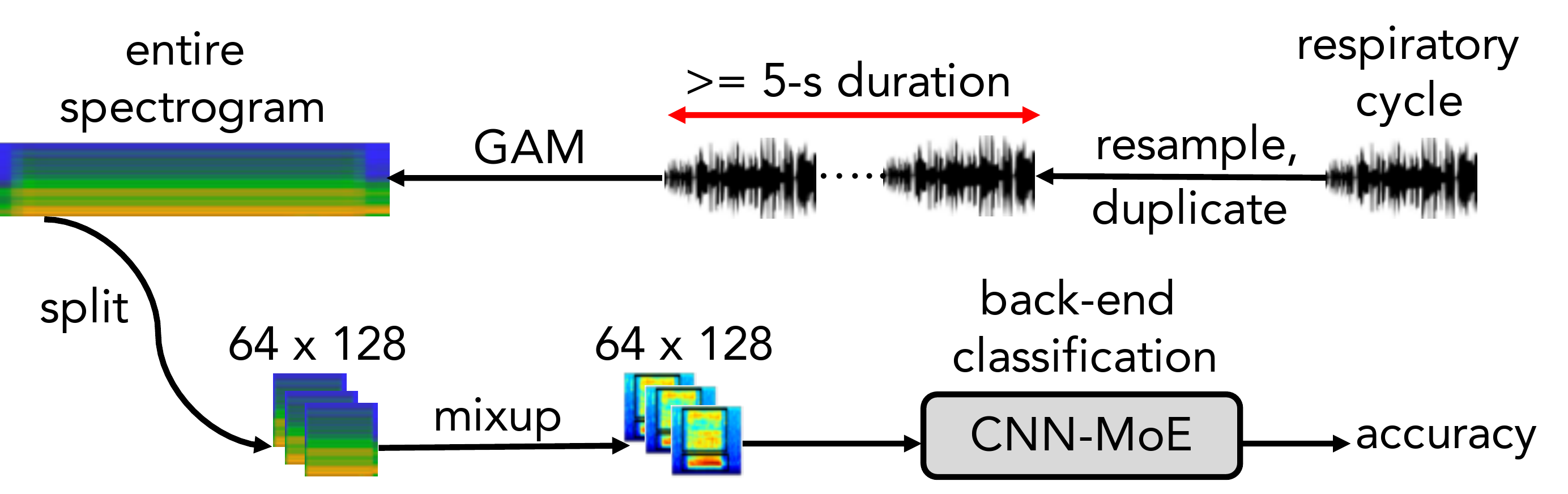}
	\caption{Baseline system proposed}
    \label{fig:A1}
\end{figure}
\begin{figure}[tb]
    \centering
    \includegraphics[width =0.8\linewidth]{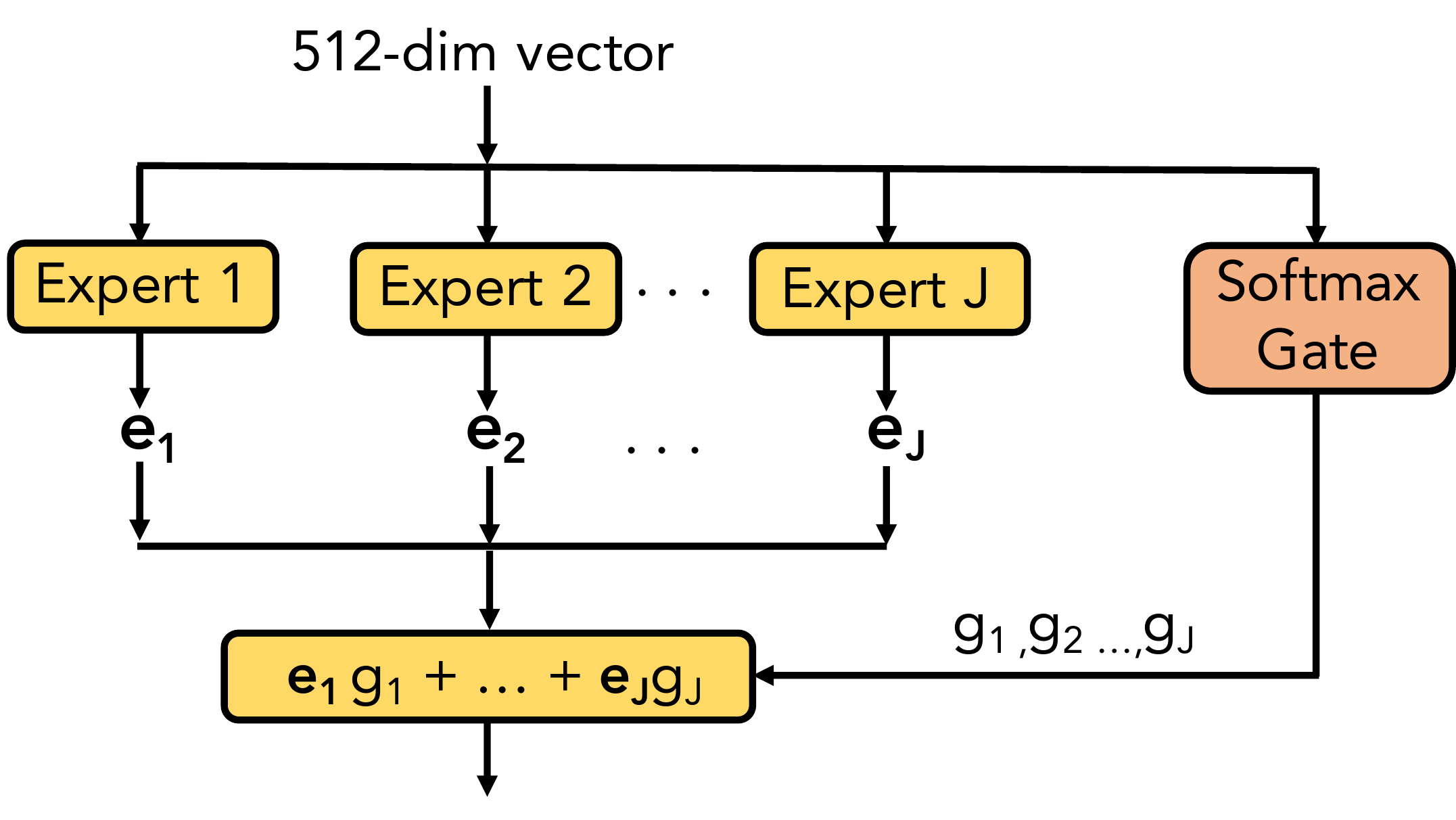}
    	\vspace{-0.3cm}
	\caption{MoE layer inside proposed CNN-MoE model}
    \label{fig:A2}
\end{figure}
\begin{table}[htb]
    \caption{CNN-MoE architecture} 
        	\vspace{-0.2cm}
    \centering
    \scalebox{0.9}{

    \begin{tabular}{l c} 
        \hline 
            \textbf{Network architecture}   &  \textbf{Output}  \\
        \hline 
         \textbf{CNN} & \\
         Input layer (image patch of $64{\times}128$)  &        \\
         Bn - Cv [$3{\times}3$] - Relu - Bn - Ap [$2{\times}2$] - Dr ($0.1\%$)      & $32{\times}64{\times}64$\\
         Bn - Cv [$3{\times}3$] - Relu - Bn - Ap [$2{\times}2$] - Dr ($0.15\%$)      & $16{\times}32{\times}128$\\
         Bn - Cv [$3{\times}3$] - Relu - Bn - Dr ($0.2\%$)      & $16{\times}32{\times}256$ \\
         Bn - Cv [$3{\times}3$] - Relu - Bn - Ap [$2{\times}2$] - Dr ($0.2\%$)       & $8{\times}16{\times}256$\\
         Bn - Cv [$3{\times}3$] - Relu - Bn  - Dr ($0.25\%$)      & $8{\times}16{\times}512$ \\
         Bn - Cv [$3{\times}3$] - Relu - Bn -  Gp - Dr ($0.25\%$) & $512$ \\           
         \hline
         \textbf{MoE} & \\
         Input layer ($512$-dimensional vectors)  &        \\
         MoE layer (10 Experts \& 1 Softmax gate) &  N       \\
         Softmax layer & N \\         
       \hline 
    \end{tabular}
    }
    \label{table:CDNN} 
\end{table}
To explore the ICBHI dataset, we firstly evaluate with the baseline system shown in Fig. \ref{fig:A1}.
This separates audio recordings into respiratory cycles based on onset and offset labels.
Since the recordings have different sample rates, we re-sample everything to 16\,kHz. Since their lengths also differ, for Task 1 we duplicate short respiratory cycles to ensure input features have a minimum length (e.g. 5\,s or longer -- this is unnecessary for Task 2 which uses entire recordings).
Each cycle is then transformed into a Gammatone spectrogram (GAM)~\cite{aud_tool} with window size=1024 samples, hop size=256, FFT length=2048 and filter number=64.
The resulting spectrogram is then split into smaller patches of size $64{\times}128$. Mixup data augmentation is applied during training~\cite{mixup_image}.
%
Back-end classification is performed by a deep learning network we call CNN-MoE which cascades a \textbf{CNN} block followed by an \textbf{MoE} network, as described in Table \ref{table:CDNN}.
The \textbf{CNN} contains sub-blocks which perform batch normalization (Bn), convolution (Cv[kernel number, kernel size]), rectified linear units (Relu), average pooling (Ap[kernel size]), global average pooling (Gp), drop out (Dr (percentage drop)), configured as shown in the upper part of Table \ref{table:CDNN}.
The \textbf{MoE} block contains a Mixture-of-Experts (MoE) layer, followed by a Softmax layer where ``N''  is the number of classes (i.e. 4 or 2 for the ICBHI Task 1 subtasks, and 3 or 2 for Task 2 subtasks, respectively).
The MoE layer, shown in Fig. \ref{fig:A2}, uses an array of different trained models (fully-connected and ReLu layers), referred to as experts~\cite{lung_moe}. 
Each expert incorporates a softmax gate (fully-connected layer followed by softmax layer) to control how each expert contributes to the output.
The final softmax layer in the {MoE} block transforms the output into a one-hot encoding.
Let  $\mathbf{e_{1}, e_{2}, \dots  e_{J}}$ be the output vectors from $J$ experts, and  $g_{1}, g_{2}, \dots , g_{J}$ be the gate network outputs. 
The predicted output of the CNN-MoE model is then,
\begin{equation}
    \label{eq:moe}
    \hat{y} = softmax \left\{ \sum_{j=1}^{J} \mathbf{e_{j}}g_{j} \right \}
\end{equation} 
During training, we use overall cross-entropy loss, 
\begin{equation}
    \label{eq:loss_func}
    Loss(\theta) = -\frac{1}{N}\sum_{i=1}^{N}y_i .log \left\{\hat{y}_{i}(\theta) \right\} + \frac{\lambda}{2}.||\theta||_{2}^{2}
\end{equation}
defined over all parameters \(\theta\), with constant \(\lambda\) set initially to $0.0001$.  \(y_{i}\) and \(\hat{y}_{i}\)  denote expected and predicted results.
The system is implemented using Tensorflow and is trained over $100$ epochs, with the Adam optimiser adjusting learning rate, on batched of size $50$.
\section{Analysis of affected factors}
Using the baseline described, we analysed feature-level factors that affect classification accuracy related to respiratory cycle length, time resolution, and network architecture.
\subsection{Cycle length analysis}
Respiratory cycles in the ICBHI dataset have lengths ranging from 0.2\,s to 16.2\,s  with 80\% of cycles being less than 5\,s, it is interesting to understand how respiratory cycle length affects classification accuracy.
Noting that our architecture duplicates short cycles to make them longer than a given minimum, we therefore adjust that minimum cycle length in our baseline from 2 to 8\,s, then retrain and evaluate Task 1 performance at each setting.

\subsection{Time resolution analysis}
This analysis uses Task 2, which classifies the type of respiratory disease over an entire audio recording. 
The baseline network operates on patches, with the horizontal dimension denoting the time span for each feature. Features are sequential, so the time span also sets the temporal resolution of the features.
To explore, we adjust patch widths to 0.6\,s, 1.2\,s, 1.8\,s, 2.4\,s and 3\,s by setting the patch dimension to be $64{\times}32$, $64{\times}64$, $64{\times}96$, $64{\times}128$, and $64{\times}160$ respectively, then retrain and evaluate performance of each.
\subsection{Network architecture analysis}
\begin{table}[tb]
    \caption{C-RNN architecture} 
        	\vspace{-0.2cm}
    \centering
    \scalebox{0.9}{

    \begin{tabular}{l c} 
        \hline 
            \textbf{Network architecture}   &  \textbf{Output}  \\
        \hline 
         Input layer (image patch of $64{\times}128$)  &        \\
         Bn - Cv [$4{\times}1$] - Relu - Bn - Ap [$2{\times}1$] - Dr ($0.1\%$)      & $32{\times}128{\times}64$\\
         Bn - Cv [$4{\times}1$] - Relu - Bn - Ap [$2{\times}1$] - Dr ($0.15\%$)      & $16{\times}128{\times}128$\\
         Bn - Cv [$4{\times}1$] - Relu - Bn - Ap [$4{\times}1$] - Dr ($0.2\%$)       & $4{\times}128{\times}256$\\
         Bn - Cv [$4{\times}1$] - Relu - Bn - Ap [$4{\times}1$] - Dr ($0.25\%$)       & $128{\times}512$\\
         bi-GRU (128 hidden states, 128 frame number) & $256{\times}512$\\
         Ap [$1{\times}512$] & $256$\\
         Fl - Relu - Dr ($0.3\%$)  &  $1024$       \\
         Fl - Relu - Dr ($0.3\%$)  &  $1024$       \\
         Fl - Softmax  &  N       \\
       \hline 
    \end{tabular}
    }
    \label{table:CRNN} 
\end{table}
%
It is known that the \textit{Crackle} and \textit{Wheeze} cycles in a respiratory recording show specific characteristics~\cite{lung_event_pro}.
In particular the former tends to have a duration of around 10\,ms, located predominantly in a frequency range of 60-2000\,Hz.
The duration of \textit{Wheeze} is longer; approximately 100 to 250\,ms and is located around 400\,Hz in frequency.
The CNN-MoE architecture proposed, while it can learn spatial features of spectrogram patches, cannot capture time-sequential features like these which exceed the duration of a patch.
This inspired us to proposed a C-RNN network combined with the CNN-MoE, to learn both spatial and time-sequential features.
The proposed C-RNN architecture is described in Table \ref{table:CRNN}.
Like the baseline CNN-MoE, patches of size $64{\times}128$ are fed into sub-blocks of Cv, Bn, ReLu, Ap and Dr.
However, the settings of these sub-blocks need to be adjusted to allow the system to learn time-sequential features.
In particular, convolutional layers with kernel size of [$4{\times}1$] are applied to learn the difference between frequency banks in each temporal frame. 
Average pooling layers (Ap [$2/4{\times}1$]) help to scale the frequency dimension of the spectrogram, but have the same time dimension of 128.
Before going through a bi-GRU layer, frequency dimension is scaled into 1, generating a sequence of 128-temporal frames.
Each temporal frame is represented by a 512-dimensional vector for the final convolutional layer.  
The bi-GRU layer generates a new sequence, increasing size from 128 to 256 due to bi-directional learning.
We apply average pooling (Ap [$1{\times}512$]) that help smooth each temporal frame, yielding a 256-dimensional vector.
These vectors are fed into three fully-connected layers with drop-out for classification.
We evaluate the two networks (CNN-MoE and C-RNN) for Task 1 and 2, and well as for their fusion \(P_{f}\), 
\begin{equation}
    \label{eq:fusion}
        P_{f} = (P_{CNN-MoE} + P_{C-RNN})/2
\end{equation} 
where \(P_{CNN-MoE}, P_{C-RNN}\) are the probability of CNN-MoE and C-RNN model, respectively. Front end feature extraction is unchanged.
\section{Experimental results}
\subsection{Cycle length comparison}
\begin{table}[htb]
	\caption{Respiratory cycle length analysis over Task 1} 
        	\vspace{-0.2cm}
    \centering
    \scalebox{0.9}{

    \begin{tabular}{c c c c c c c} 
        \hline 
	    \textbf{Classes}  &\textbf{Cyc. Len.}    &\textbf{Spec.}   &\textbf{Sen.}   &\textbf{ICBHI Score}  \\
        \hline 
	    4                 &2\,s                    &0.87             &0.70            &0.78  \\
	    4                 &3\,s                    &0.86             &0.71            &0.78  \\
	    4                 &4\,s                    &0.87             &0.69            &0.78  \\
	    4                 &5\,s                    &0.90             &0.68            &0.79  \\
	    4                 &\textbf{6\,s}                    &\textbf{0.90}    &\textbf{0.70}   &\textbf{0.80}  \\
	    4                 &7\,s                    &0.86             &0.72            &0.79  \\
	    4                 &8\,s                    &0.89             &0.69            &0.79  \\
        \hline 
        \hline 
	    2                 &2\,s                    &0.87             &0.80            &0.83  \\
	    2                 &3\,s                    &0.86             &0.81            &0.84  \\
	    2                 &4\,s                    &0.87             &0.79            &0.83  \\
	    2                 &5\,s                    &0.90             &0.78            &0.84  \\
	    2                 &\textbf{6\,s}                    &\textbf{0.90}    &\textbf{0.80}   &\textbf{0.85}  \\
	    2                 &7\,s                    &0.86             &0.82            &0.84  \\
	    2                 &8\,s                    &0.89             &0.79            &0.83  \\
       \hline 
    \end{tabular}
    }
    \label{table:cyc_ana} 
\end{table}
%
%
%
The Table \ref{table:cyc_ana} reports Task 1 results for different cycle lengths.
Both 4-class (\textit{Crackle, Wheeze, Both} and \textit{Normal}) and 2-class (\textit{Normal} and \textit{Others}) scores improve as minimum cycle length exceeds 4\,s. 
The best ICBHI scores are 0.80 and 0.85  for 4- and 2-class sub-tasks respectively. Interestingly, both are achieved with a 6\,s cycle length.

\subsection{Time resolution comparison}
The results, presented in Table \ref{table:time_ana}, show that patch size has quite a large influence on ICBHI score.
The best results of 0.91 and 0.92 for the 3-class sub-task (\textit{Healthy, Chronic diseases} and \textit{Non-chronic diseases}) and 2-class sub-task (\textit{Healthy} or \textit{Unhealthy}) respectively, occur with a patch length of 2.4\,s.
Again it is interesting that both tasks agree on the preferred time resolution.
\subsection{Deep learning model comparison}
%
\begin{table}[htb]
	\caption{Time resolution analysis over Task 2} 
        	\vspace{-0.2cm}
    \centering
    \scalebox{0.9}{

    \begin{tabular}{c c c c c c c} 
        \hline 
	    \textbf{Classes}  &\textbf{Time len.}    &\textbf{Spec.}     &\textbf{Sen.}             &\textbf{ICBHI Score}  \\
        \hline 
	    3                 &0.6\,s                  &0.57               &0.97                      &0.77  \\
	    3                 &1.2\,s                  &0.57               &0.96                      &0.77  \\
	    3                 &1.8\,s                  &0.71               &0.97                      &0.84  \\
	    3                 &\textbf{2.4\,s}                 &\textbf{0.86}      &\textbf{0.96}             &\textbf{0.91}  \\
	    3                 &3.0\,s                  &0.71              &0.97                      &0.84  \\
	    3                 &3.6\,s                  &0.71              &0.95                      &0.83  \\

        \hline 
        \hline
	    2                 &0.6\,s                  &0.57          &0.99                 &0.78  \\
	    2                 &1.2\,s                  &0.57          &1.00                 &0.79  \\
	    2                 &1.8\,s                  &0.71          &1.00                 &0.85  \\
	    2                 &\textbf{2.4\,s}                 &\textbf{0.86} &\textbf{0.99}        &\textbf{0.92}  \\
	    2                 &3.0\,s                  &0.71          &0.99                &0.85  \\
	    3                 &3.6\,s                  &0.71          &0.98             &0.85  \\

        \hline 
    \end{tabular}
    }
    \label{table:time_ana} 
\end{table}
\begin{table}[htb]
	\caption{Deep learning model comparison} 
        	\vspace{-0.2cm}
    \centering
    \scalebox{0.9}{

    \begin{tabular}{c c c c c c c} 
        \hline 
	    \textbf{Model}  &\textbf{Task}    &\textbf{Spec.}     &\textbf{Sen.}             &\textbf{ICBHI Score}  \\
        \hline 
	     1, 4-class       &CNN-MoE                     &0.90               &0.68                      &0.79  \\
	     1, 4-class       &CRNN                        &0.86               &0.72                      &0.79  \\
	     1, 4-class       &Ensemble                    &\textbf{0.89}               &\textbf{0.72}                      &\textbf{0.80}  \\
        \hline 
	     1, 2-class       &CNN-MoE                     &0.90               &0.78                      &0.84  \\
	     1, 2-class       &CRNN                        &0.86               &0.83                      &0.85  \\
	     1, 2-class       &Ensemble                    &\textbf{0.89}               &\textbf{0.82}                      &\textbf{0.85}  \\
        \hline 
        \hline
	     2, 3-class       &CNN-MoE                     &\textbf{0.86}               &\textbf{0.96}                      &\textbf{0.91}  \\
	     2, 3-class       &CRNN                        &0.57               &0.94                      &0.76  \\
	     2, 3-class       &Ensemble                    &0.71               &0.95                      &0.83  \\
        \hline 
	     2, 2-class       &CNN-MoE                     &\textbf{0.86}          &\textbf{0.99}                 &\textbf{0.92}  \\
	     2, 2-class       &CRNN                        &0.57          &0.98                 &0.77  \\
	     2, 2-class       &Ensemble                    &0.71          &0.99                 &0.85  \\
        \hline 
    \end{tabular}
    }
    \label{table:model_ana} 
\end{table}
%

%
Table \ref{table:model_ana} reports classification accuracy for Task 1 and 2 with different back-end classifiers, namely the CNN-MoE (deep learning model baseline),  C-RNN, and their ensemble.
For Task 1, CNN-MoE and C-RNN have similar performance, achieving 0.79 and around 0.84 for 4-class and 2-class sub-tasks, respectively. 
An ensemble helps to enhance the performance, but the improvement is small.
For Task 2, the CNN-MoE outperforms C-RNN over every sub-task, and the ensemble fails to improve performance.
The best results on CNN-MoE are 0.91 and 0.92 for 3-class and 2-class sub-tasks, respectively. 
\subsection{Compared to the state of the art}
\begin{table}[htb]
    \caption{Comparison against state-of-the-art systems} 
        	\vspace{-0.2cm}
    \centering
    \scalebox{0.9}{

    \begin{tabular}{l l c c c c} 
        \hline 
	    \textbf{Task}   &\textbf{Method}    &\textbf{Spec.}   &\textbf{Sen.}   &\textbf{ICBHI Score}  \\
        \hline 
	    1, 4-class      &Boosted Tree~\cite{lung_tree_18}           &0.78             &0.21            &0.49  \\
	    1, 4-class      &CNN~\cite{lung_cnn_02}                    &0.77             &0.45            &0.61  \\
	    1, 4-class      &MNRNN~\cite{lung_rnn_02}                  &0.74             &0.56            &0.65  \\
	    1, 4-class      &PCA variance~\cite{lung_wavelet_01}           &0.83             &0.55            &0.69  \\
	    1, 4-class      &LSTM~\cite{lung_rnn_01}                   &0.85             &0.62            &0.74  \\
	    1, 4-class      &\textbf{Our system (ensemble)}  &\textbf{0.86}                   &\textbf{0.73}                  &\textbf{0.80}    \\
       \hline 
	    1, 2-class      &LSTM~\cite{lung_rnn_01}                   &-                &-               &0.81  \\
	    1, 2-class      &\textbf{Our system (ensemble)}  &\textbf{0.86}                 &\textbf{0.85}                  &\textbf{0.86}    \\
       \hline 
       \hline
	    2, 3-class      &CNN~\cite{lung_cnn_02}                    &0.76             &0.89            &0.83  \\
	    2, 3-class      &LSTM~\cite{lung_rnn_01}                   &0.82             &\textbf{0.98}            &\textbf{0.90}   \\
	    2, 3-class      &\textbf{Our system (CNN-MoE)}            &\textbf{0.83}                 &0.96                &\textbf{0.90}      \\
       \hline 
	    2, 2-class      &CNN~\cite{lung_cnn_02}                     &0.78             &0.97            &0.88  \\
	    2, 2-class      &LSTM~\cite{lung_rnn_01}                   &0.82             &\textbf{0.99}    &\textbf{0.91} \\
	    2, 2-class      &\textbf{Our systen (CNN-MoE)}                      &\textbf{0.83}   &\textbf{0.99}    &\textbf{0.91} \\
       \hline 
    \end{tabular}
    }
    \label{table:comp_sta} 
\end{table}
%
From the experimental analysis results, we propose separate network configurations for ICBHI challenge Tasks 1 and 2, although both share the same front-end feature extraction settings.
The Task 1 system adopts the baseline architecture, but sets a 6\,s minimum cycle length and uses an ensemble of CNN-MoE and C-RNN for classification.
Task 2 uses the baseline architecture with the CNN-MoE classifier.

Table \ref{table:comp_sta}(top) compares the proposed Task 1 system against the state of the art, demonstrating the highest accuracy of 0.80 and 0.86 for the 4-class and 2-class subtasks, respectively.
Task 2 results (Table \ref{table:comp_sta}, bottom) reveal an accuracy of 0.91 and 0.90 for the 3-class and 2-class subtasks respectively, matching \cite{lung_rnn_01} and outperforming the other publications.
The fact that the same feature extraction process, range of settings and baseline architecture (with the exception of final classifier layer), performs well on all tasks serves as an indicator of system robustness.
\section{Conclusion}
This paper has presented an exploration of deep learning models for detecting respiratory disease from auditory recordings.
By conducting intensive experiments over the ICBHI  dataset, we propose a deep learning framework for four challenge tasks of respiratory sound classification.
The proposed system is shown to outperform the state of the art on two tasks, and match it on two, validating this application of deep learning for  early diagnosis of respiratory disease.

\addtolength{\textheight}{-12cm}   

\bibliographystyle{IEEEbib}
\bibliography{refs}


\end{document}